\documentclass[11pt]{article}
\usepackage{graphicx}
 
\newcommand{\BABARPubYear}    {04}

\newcommand{\BABARConfNumber} {50}
\newcommand{\SLACPubNumber} {10646}
\newcommand{\LANLNumber} {0000000}

\input babarsym
   
\long\def\inst#1{\par\nobreak\kern 4pt\nobreak
    {\it #1}\par\vskip 10pt plus 3pt minus 3pt}
 
\newcommand{\beq}       {\begin{equation}}
\newcommand{\eeq}       {\end{equation}}
\newcommand{\etal}      {{\it et~al.}}

\newcommand{\bsnunu}       {\mbox{$\b \to \s \nunub$}}
\newcommand{\bdnunu}       {\mbox{$\b \to \d \nunub$}}

\newcommand{\bsll}         {\mbox{$\b \to \s \ellell$}}
\newcommand{\Bknunu}       {\mbox{$\Bp \to \Kp \nunub$}}
\newcommand{\Bpinunu}       {\mbox{$\Bp \to \pip \nunub$}}

\newcommand{\tagB}        {\mbox{$B^-$}}
\newcommand{\tagBlep}        {\mbox{$B^-_{\rm sl}$}}
\newcommand{\tagBhad}        {\mbox{$B^-_{\rm had}$}}

\newcommand{\ksigeffLep}   {\mbox{$\varepsilon_{K}=(0.115 \pm 0.009)\%$}}

\newcommand{\ksigeffHad}   {\mbox{$\varepsilon_{K}=(0.055 \pm 0.005)\%$}}
\newcommand{\pisigeffHad}  {\mbox{$\varepsilon_{\pi}=(0.065 \pm 0.006)\%$}}

\newcommand{\Xhad}        {\mbox{$X^-_{\rm had}$}}
\newcommand{\BptoDlnu}      {\mbox{$\Bp \to \Dzb \ellp \nu$}}

\newcommand{\eeqq}      {\mbox{$e^{+}e^{-} \to q \overline {q}$}}

\newcommand{\Dsl}        {\ensuremath{D}\ell\xspace}

\newcommand{\Ebeam}       {\mbox{$E_{\rm beam}$}}
\newcommand{\Eextra}       {\mbox{$E_{\rm extra}$}}

\newcommand{\costht}       {\mbox{$|\cos \theta_T|$}}

\newcommand{\cosBDl}       {\mbox{$\cos \theta_{B,D\ell}$}}

\newcommand{\samplelumi}  {\mbox{$82 \invfb$}}
\newcommand{\samplelumioffpeak}  {\mbox{$9.6 \invfb$}}
                                                                                                                                                            
\begin{document}
{\pagestyle{empty}
 
\begin{flushright}
\babar-CONF-\BABARPubYear/\BABARConfNumber \\
SLAC-PUB-\SLACPubNumber \\
hep-ex/\LANLNumber \\
July 2004 \\
\end{flushright}
 
\par\vskip 5cm
 
\begin{center}
\Large \bf
 A search for the decay {$\mathbf \Bknunu$}
\end{center}
\bigskip
                                                                                                                                                            
\begin{center}
\large The \babar\ Collaboration\\
\mbox{ }\\
\today
\end{center}
\bigskip \bigskip
                                                                                                                                                            
\begin{center}
\large \bf Abstract
\end{center}
We present a search for the rare flavor-changing neutral-current decay $\Bknunu$ based on data sample
of $\samplelumi$ collected with the \babar\ detector at the PEP-II B-factory.
Signal events are selected by examining the properties of the system recoiling against either a
reconstructed hadronic or semileptonic charged $B$ decay.  Using these two independent
samples, a combined limit of ${\mathcal B}(\Bknunu)<5.2 \times 10^{-5}$ is obtained at the $90\%$ confidence level.
In addition, by modifying the particle identification criteria, a limit of ${\mathcal B}(\Bpinunu)<1.0 \times 10^{-4}$
obtained using the hadronic $B$ reconstruction.  All results are preliminary.
\vfill
\begin{center}

Submitted to the 32$^{\rm nd}$ International Conference on High-Energy Physics, ICHEP 04,\\
16 August---22 August 2004, Beijing, China
                                                                                                                                                            
\end{center}
                                                                                                                                                            
\vspace{1.0cm}
\begin{center}
{\em Stanford Linear Accelerator Center, Stanford University,
Stanford, CA 94309} \\ \vspace{0.1cm}\hrule\vspace{0.1cm}
Work supported in part by Department of Energy contract DE-AC03-76SF00515.
\end{center}
                                                                                                                                                            
\newpage
} 
                                                                                                                                                            
%
%
\begin{center}
\small

The \babar\ Collaboration,
\bigskip

%
B.~Aubert,
R.~Barate,
D.~Boutigny,
F.~Couderc,
J.-M.~Gaillard,
A.~Hicheur,
Y.~Karyotakis,
J.~P.~Lees,
V.~Tisserand,
A.~Zghiche
\inst{Laboratoire de Physique des Particules, F-74941 Annecy-le-Vieux, France }
% ...

\end{center}\newpage





\noindent Flavor-changing neutral-current transitions such 
as $\bsnunu$ and $\bdnunu$ occur in the Standard Model (SM) via one-loop
box or electroweak penguin diagrams with virtual heavy particles in the loops.  Therefore they are expected  
to be highly suppressed.  Because heavy non-SM particles could contribute additional loop diagrams,
various new physics scenarios can potentially lead to significant 
enhancements in the observed rates~\cite{ref:grossman}.
Theoretical uncertainties on $\bsnunu$ are much smaller than the corresponding $\bsll$ modes due to the absence of
a photonic penguin contribution and hadronic long distance effects~\cite{ref:faessler}.
The SM $\Bknunu$ branching fraction has been estimated to be
$(3.8^{+1.2}_{-0.6}) \times 10^{-6}$~\cite{ref:faessler,ref:buchalla}, while
the most stringent published experimental limit is
$\mathcal{B}(\Bknunu) < 2.4 \times 10^{-4}$ at the $90\%$ confidence level (C.L.)~\cite{ref:cleo}.
$\bdnunu$ processes are additionally suppressed relative to $\bsnunu$ by the 
ratio of the Cabibbo-Kobayashi-Maskawa matrix elements 
$|V_{td}|^2/|V_{ts}|^2 $~\cite{ref:aliev}.
 
In this work we report the results of a search for the exclusive decay mode $\Bknunu$.
By modifying the particle identification (PID) criteria used in the search,
we additionally obtain a limit on the related decay $\Bpinunu$.

The data used in this analysis were collected with the 
\babar\ detector~\cite{ref:babar} at the
PEP-II asymmetric-energy $e^{+}e^{-}$ storage ring.  
The results are based on a data sample
of 88.9 million $\B\Bbar$ events, corresponding to an integrated luminosity
of $\samplelumi$ collected at the $\Y4S$ resonance. 
 An additional sample of $\samplelumioffpeak$ was collected at a center-of-mass
 (CM) energy approximately 40~MeV below $\B\Bbar$ threshold 
which is used to study continuum events, $\eeqq~(q = u,d,s $ and $c)$. 
Charged particle tracking and $\dedx$ measurement for particle identification (PID) are 
provided by a five-layer double-sided silicon vertex tracker
and a 40-layer drift chamber contained within the magnetic field of a 
1.5${-}$T superconducting solenoid.
Charged $K{-}\pi$ PID separation of greater than $3\sigma$, over the momentum range 
of interest for this analysis, is provided by a ring-imaging Cherenkov 
detector (DIRC).
  The energies of neutral particles are measured by an electromagnetic 
calorimeter (EMC) consisting of 6580 CsI(Tl) crystals.
The magnetic flux return of the solenoid is instrumented with 
resistive plate chambers in order to provide muon identification.
A full \babar\ detector  Monte Carlo (MC) simulation based on 
{\tt GEANT4}~\cite{ref:geant4} is used to evaluate 
signal efficiencies and to identify and study background sources.
Charge conjugate modes are implied throughout this paper and all 
kinematic quantities are expressed in the CM
frame (i.e. the $\Y4S$ rest frame) unless otherwise specified.  

The presence of two neutrinos in the final state precludes the direct
reconstruction of the $\Bknunu$ signal mode. Instead, 
the $\tagB$ meson from an $\Y4S \to B^+B^-$ event is reconstructed
in one of many semileptonic or hadronic decay modes, 
then all remaining charged and neutral particles in that event are examined
under the assumption that they are attributable to the decay
of the accompanying $B$. 

The $\tagB$ reconstruction proceeds by combining a $\Dz$ candidate
with either a single identified charged lepton or a combination, $\Xhad$, 
of charged and neutral hadrons.  The resulting semileptonic and hadronic charged $B$ 
samples are referred to as $\tagBlep$ and $\tagBhad$ throughout this paper.
$\Dz$ candidates are reconstructed by selecting combinations of
 identified pions and kaons which yield an invariant mass within 
approximately $3\sigma$ of the expected $\Dz$ mass in the modes $\Km\pip$, $\Km\pip\piz$ and $\Km\pip\pip\pim$.
For $\tagBhad$ reconstruction, $\Dz \to K^{0}_{s}\pip\pim$ is also used.

Photon candidates are obtained from EMC clusters with laboratory-frame energy greater than $30$~MeV and no 
associated charged track. 
Photon pairs which combine to yield $\gamma \gamma$
 invariant mass between 115~MeV$/c^2$ and 150~MeV$/c^2$ and total energy greater than $200$~MeV are
considered to be $\piz$ candidates.

$\tagBlep$ candidates are reconstructed by combining a
$\Dz$ candidate having a momentum $p_{D^{0}}>0.5$~GeV$/c$ with a lepton candidate of momentum $p_{\ell}>1.35$ $\gevc$ 
that satisfies either electron or muon identification criteria.  
The invariant mass, $m_{{\Dsl}}$, of the $\Dz\ell$ candidate is required to be greater than $3.0$~GeV$/c^2$.
$\tagBlep$ candidates are selected using the quantity $\cosBDl$, which represents
the cosine of the angle between the inferred direction of the reconstructed $\tagBlep$ and that of the
lepton--$\Dz$ combination, described by the four vector, $(E_{{\Dsl}}, \bf{p}_{\it{\Dsl}})$. 
Under the assumption that the neutrino is the only missing particle, 
\beq
{\cosBDl} \equiv \frac{2\,\Ebeam \cdot E_{{\Dsl}} -m^2_{{\B}} - m^2_{{\Dsl}}}
{2\, |{\bf{p}}_{\it{\Dsl}}|  \cdot \sqrt{E^2_{\rm beam}-m^2_B}   }  
\eeq
where $m_B$ is the nominal $B$ meson mass
and $\Ebeam$ and $\sqrt{E^2_{\rm beam}-m^2_B}$ are the expected $B$ meson energy 
and momentum, respectively.  
Combinatorial backgrounds can produce values, $|\cosBDl|>1$. 
In order to maintain efficiency for $B^- \to D^{*0} \ell^- \bar{\nu}$ decays
in which a $\piz$ or photon has not been reconstructed as part of the  $D\ell$ combination,  
we retain events in the interval $-2.5<${\cosBDl}$<1.1$. However events are vetoed if a charged $\pi$ 
consistent with a $D^{*+}$ transition is identified.
If more than one $D\ell$ candidate is reconstructed in a given event,
the candidate with the smallest $|${\cosBDl}$|$ is retained.

\begin{figure}[!htb]
\begin{center}
{\includegraphics[height=4.1cm]{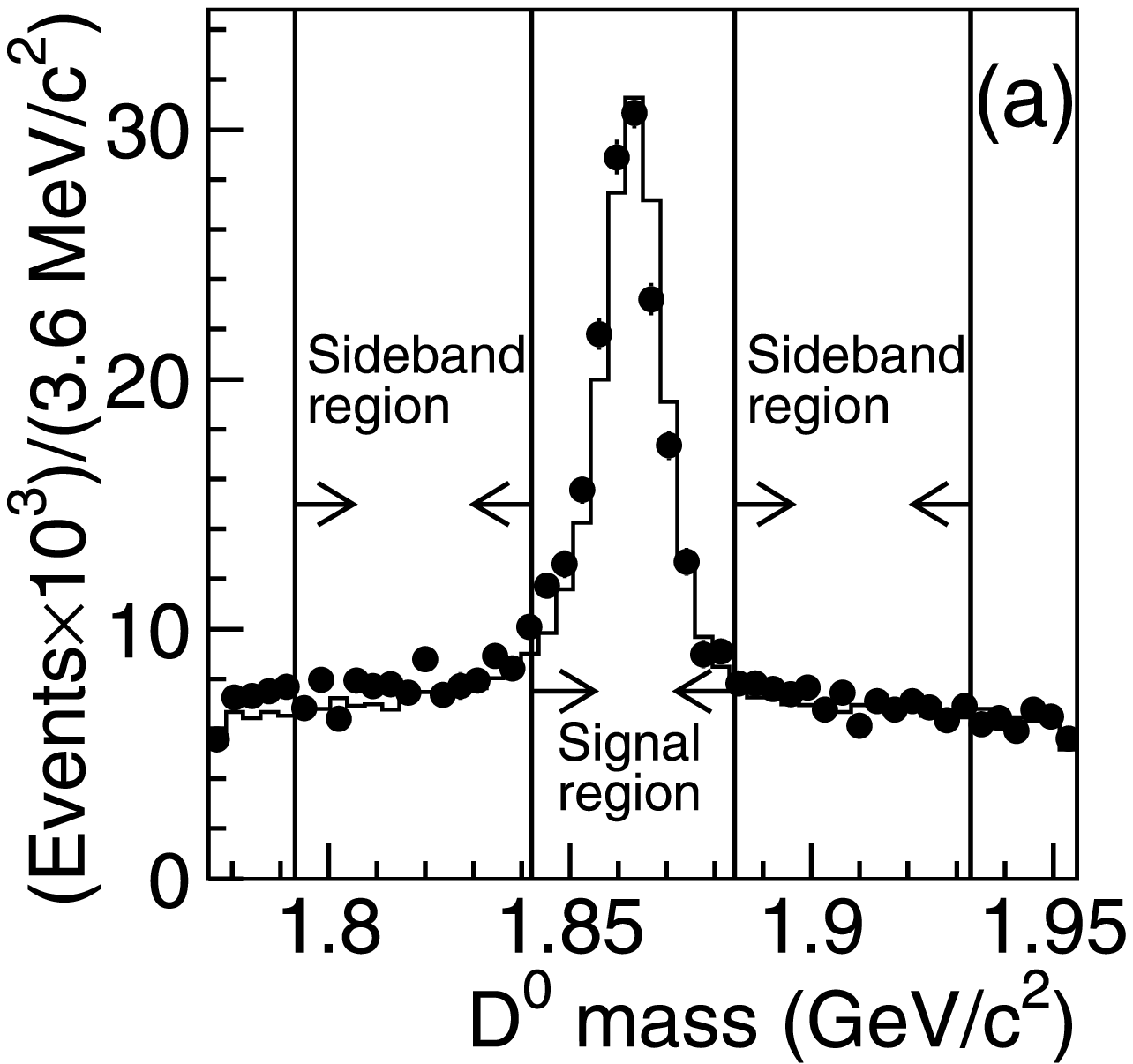}}
{\includegraphics[height=4.1cm]{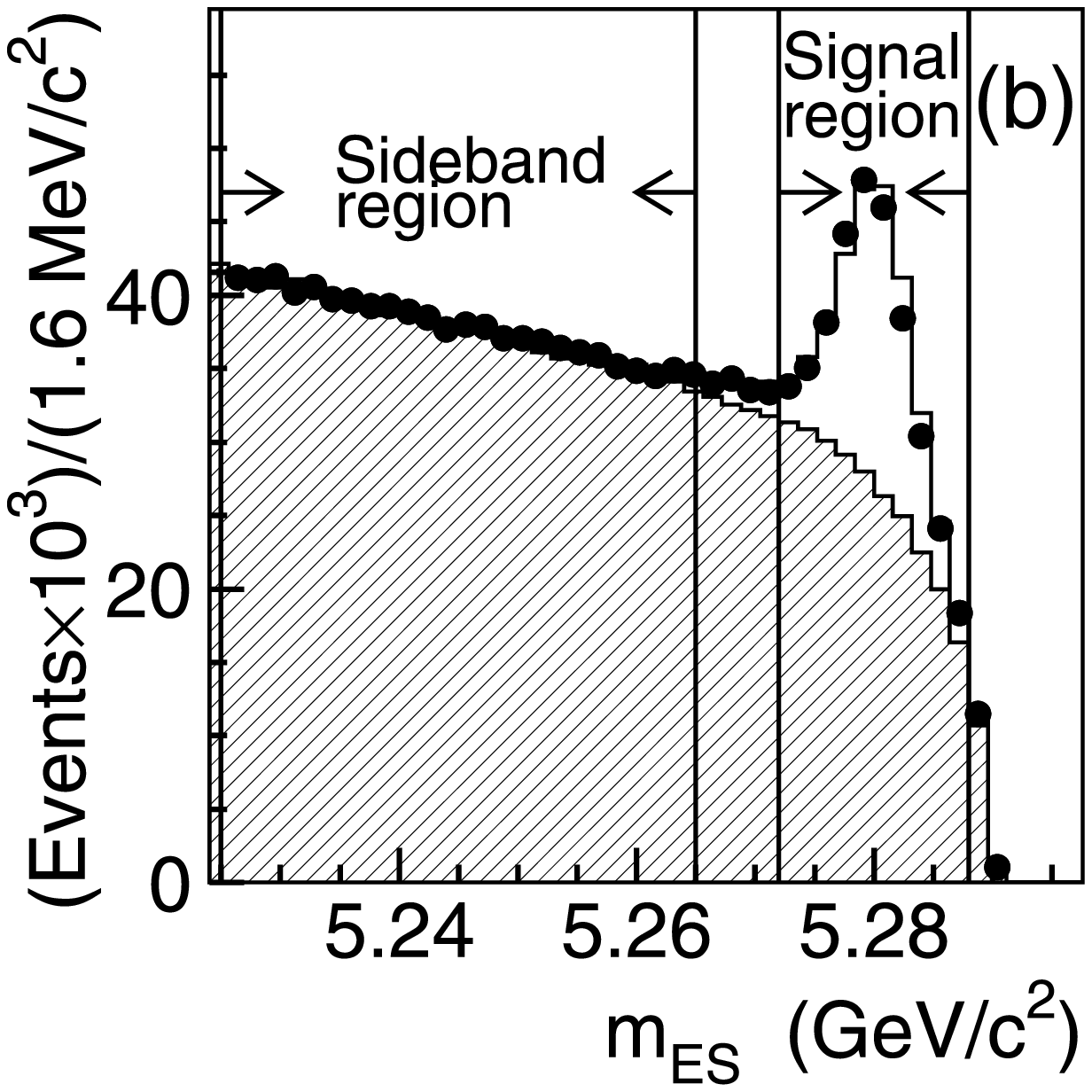}}
\caption{(a)  The $\Dz$ mass distribution for $\Dz \to \Km\pip$ decays used for $\tagBlep$ reconstruction. 
Data are shown as points and the total background MC is shown as a solid histogram. 
(b) The $\mes$ distribution of $\tagBhad$ events for data (points) 
and $B\Bbar$ MC (solid histogram).
Continuum background has been subtracted from the on-resonance data using off-resonance data and  
the hatched histogram represents the estimated combinatorial background from $B\Bbar$ decays.}
\label{fig:mes2}
\end{center}
\end{figure}

Reconstructed $\tagBhad$ decays are obtained by combining a reconstructed $\Dz$ candidate with
a hadronic system $\Xhad$ composed of up to five mesons ($\pi^\pm, K^\pm$ and $\piz$), including up to
two $\piz$ candidates.  We define the kinematic variables  $m_{ES} \equiv \sqrt{{E^2_{\rm beam}}-{\bf{p}}^{2}_{\it B}}$ and $\Delta E \equiv E_{B}-\Ebeam$ 
where $\bf{p}_{\it B}$ and $E_{B}$ are the momentum and the energy of the
$\tagBhad$ candidate. 
The $\Xhad$ system is selected by requiring that
 the resulting $\tagBhad$ candidate lies within $-1.8<\Delta E < 0.6$~GeV.
  If multiple $\tagBhad$ candidates are identified in an event, 
only the one with $\Delta E$ closest to zero is retained. 
 The $\mes$ distribution of reconstructed $\tagBhad$ candidates is shown in Fig.~\ref{fig:mes2}b. 
$\tagBhad$ candidates in the signal region,  5.272$<\mes<$5.288~GeV$/c^2$, are used for the $\Bknunu$ signal selection. 
Candidates in the sideband region, 5.225$<\mes<$5.265~GeV$/c^2$,  are retained for background studies.

Combinatorial backgrounds from continuum 
 events are reduced in both the $\tagBlep$ and $\tagBhad$ samples by requiring
  $\costht <0.8$, where $\theta_T$ is the angle between the thrust axes
 defined by the $\tagBlep$ or $\tagBhad$ daughter particles, and by all other tracks and clusters in the event.
Continuum events peak at  $\costht = 1$, while the distribution is approximately flat for $\BB$ events.
Backgrounds from QED processes are strongly suppressed by the $\tagB$ reconstruction procedures and are negligible in this analysis. 

The $\tagB$ reconstruction efficiency for $\Bknunu$ signal events is determined from signal MC simulation after validating the
yield from $B^+B^-$ MC simulation against data.  This procedure compensates for differences in the $\tagBhad$
reconstruction efficiency in the low-multiplicity environment of $\Bknunu$ events compared with the generic $B^+B^-$ environment. 
The $\tagB$ reconstruction efficiency in MC is additionally validated by comparing the yield of events in which a $\BptoDlnu$ has 
been reconstructed in addition to the $\tagBhad$ or $\tagBlep$.  

The $\tagBlep$ and $\tagBhad$ reconstruction procedures result in raw yields of approximately
5800 $\tagBlep$/$\invfb$ and 2200 $\tagBhad$/$\invfb$.
 Relative systematic uncertainties of $4.5\%~(7\%)$ are estimated for the 
overall $\tagBlep~(\tagBhad)$ yields. 

Events that contain a reconstructed  $\tagB$ are examined for
evidence of a  $\Bknunu$ decay.  Tracks and EMC clusters
not already utilized for the $\tagB$ reconstruction are assumed to be the daughters of 
the signal candidate $B$ decay.
Signal candidate events are required to possess exactly one additional charged track with charge opposite that of the 
reconstructed $\tagB$.  The track is additionally required to have 
 momentum $p_{K} > 1.25$~GeV$/c$ and to satisfy $K$ PID criteria. 

\begin{figure}[!htb]
\begin{center}
\parbox{4.0cm}
{\includegraphics[width=4.0cm,height=3.5cm]{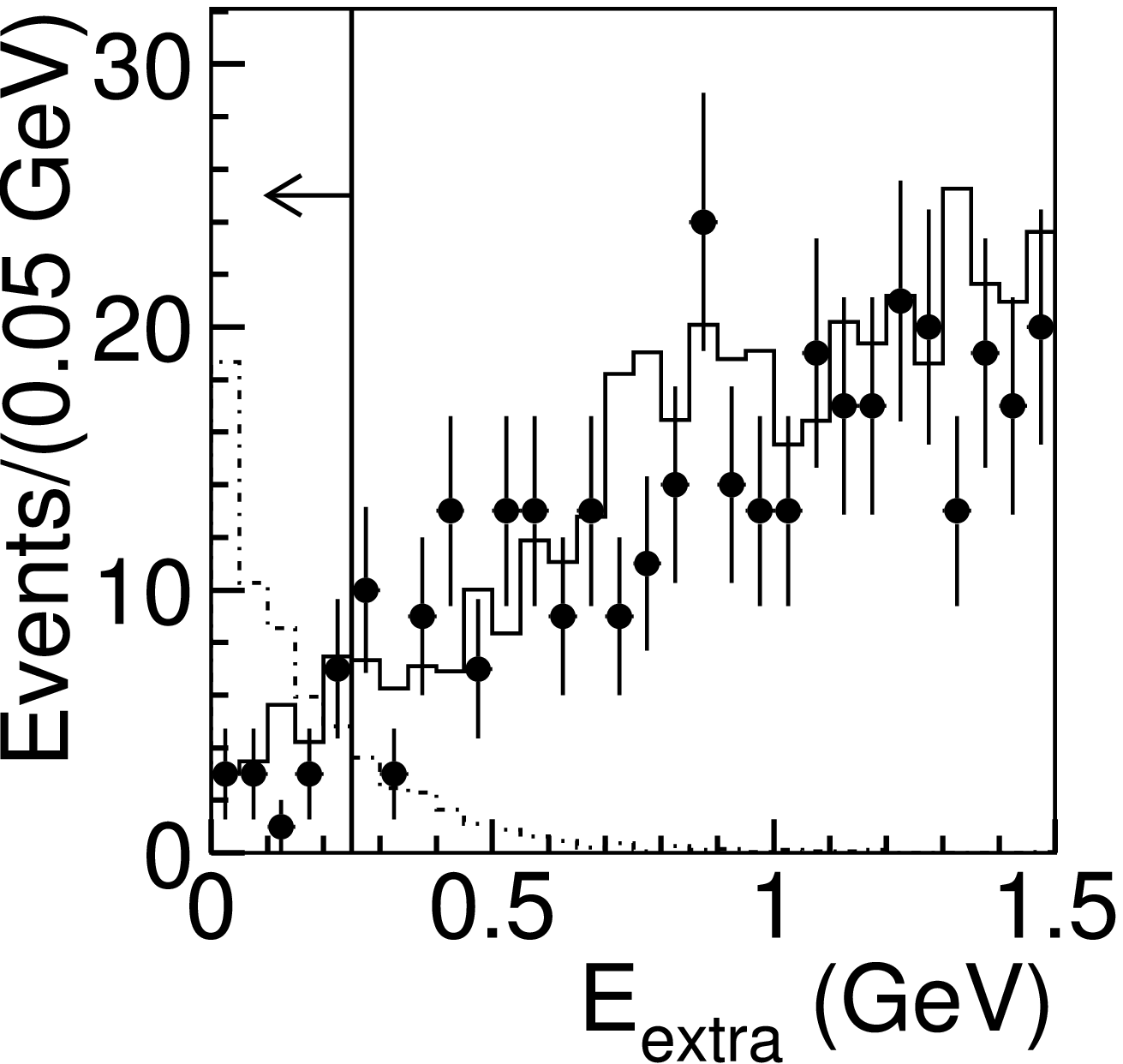}}
\parbox{4.0cm}
{\includegraphics[width=4.0cm,height=3.5cm]{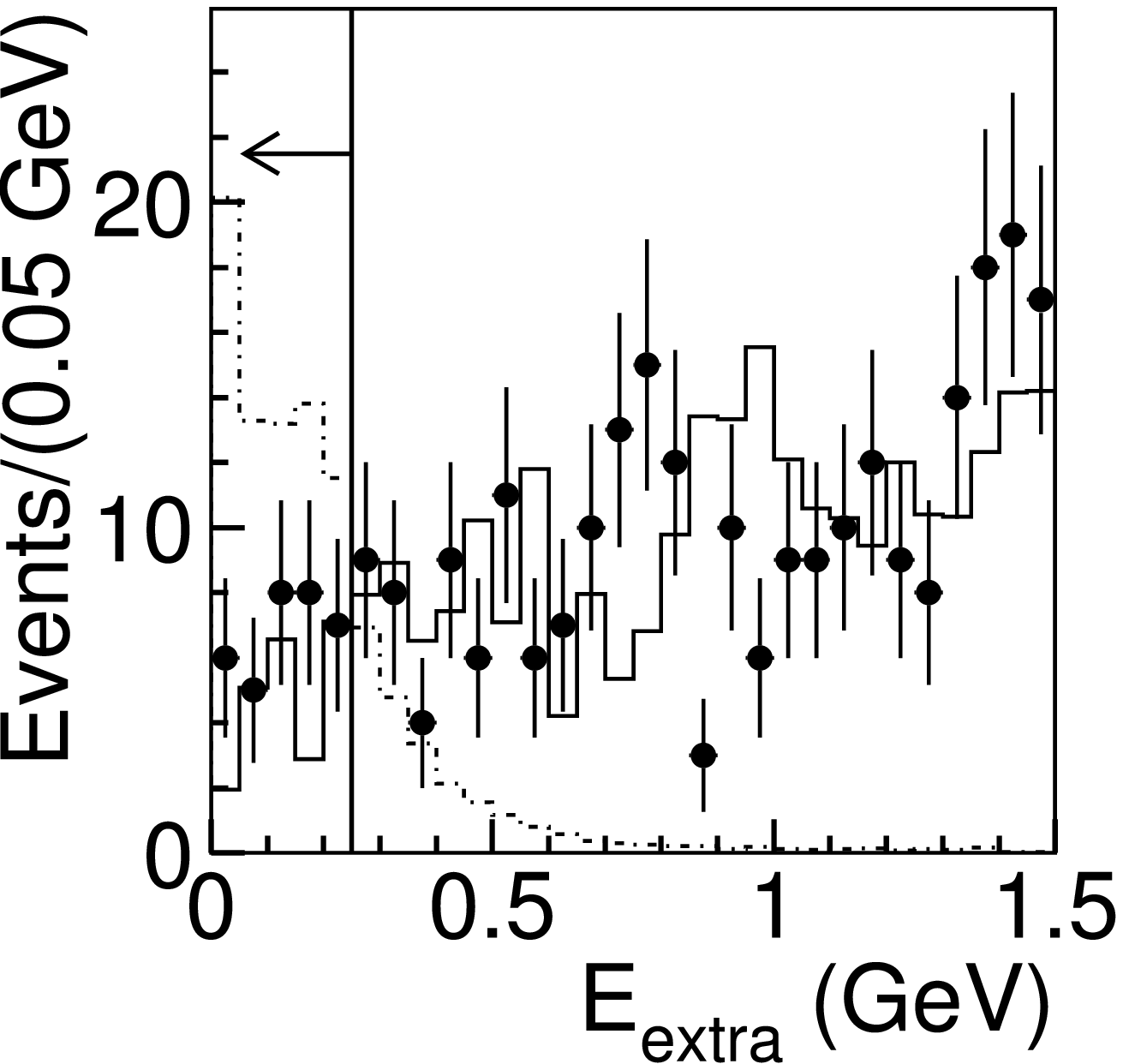}}
\caption{ The $\Eextra$ distribution for $\Bknunu$ (left) 
$\tagBhad$ (right) $\tagBlep$ events. Events are 
required to have a reconstructed $\tagB$ and exactly one additional track 
which has been identified as a kaon.  No other signal selection cuts have been applied.
The data and background MC samples are represented by the 
points with error bars and solid histograms, respectively. 
The dotted line indicates the $\Bknunu$ MC prediction with arbitrary normalization. }
\label{fig:eextra}
\end{center}
\end{figure}

In addition to this track, $\Bknunu$ events contain an average of \mbox{$\sim 200$~MeV} of EMC energy 
from hadronic shower fragments, photons from unreconstructed $D^* \to \Dz \gamma/\piz$ transitions 
in the $\tagB$ candidate, and beam-related background photons.  
The total calorimeter energy attributed to the signal decay, $\Eextra$, is computed by
 summing all EMC clusters
that are not associated either with the decay daughters of the $\tagB$ or with the signal track. 
Signal events are required to have $\Eextra <250$~MeV.
The $\Eextra$ distributions are shown in Fig.~\ref{fig:eextra} for $\tagBlep$ and 
$\tagBhad$ events with one additional track which has been identified as a kaon.   
The $\tagBhad$ analysis additionally
requires that there are six or fewer clusters contributing to $\Eextra$, and that no pair of these clusters
 can be combined to form a $\piz$ candidate.

 The total $\Bknunu$ signal selection efficiencies, including the $\tagB$ reconstruction, are estimated to be 
$\ksigeffLep$ for $\tagBlep$ and $\ksigeffHad$ for $\tagBhad$ events.  
The quoted errors are the quadratic sum of statistical and systematic uncertainties.
 Theoretical uncertainties in the $K^\pm$ energy spectrum result in a $1.3\%$ uncertainty on 
 the signal efficiency.  This uncertainty is evaluated by
comparing the $p_K$ spectrum of $\Bknunu$ MC events generated with a
phase-space model with the models given in~\cite{ref:buchalla,ref:faessler}.
Additional systematic uncertainties associated with the $\Bknunu$ signal
candidate efficiencies include the single track efficiency ($1.3\%$), PID ($2\%$)
and EMC energy modeling ($3.8\%$ for $\tagBlep$ and $2.3\%$ for $\tagBhad$).
The EMC energy modeling systematic is determined by 
evaluating the effect of varying the MC $\Eextra$ distribution within a range representing the observed
level of agreement with data in events with a reconstructed $\BptoDlnu$ (for the $\tagBlep$ sample)
and in samples containing two or three additional tracks (for the $\tagBhad$ sample).

Background events can arise either from $\BzBzb$ or continuum events in which 
the $\tagB$ candidate is constructed from a random combination of particles, 
or peaking background events in which the accompanying $\tagB$
(or in the case of $\tagBlep$, at least the $\Dz$) has been correctly reconstructed.

\begin{figure}[!htb]
\begin{center}
\parbox{4.0cm}
{\includegraphics[width=4.0cm,height=3.5cm]{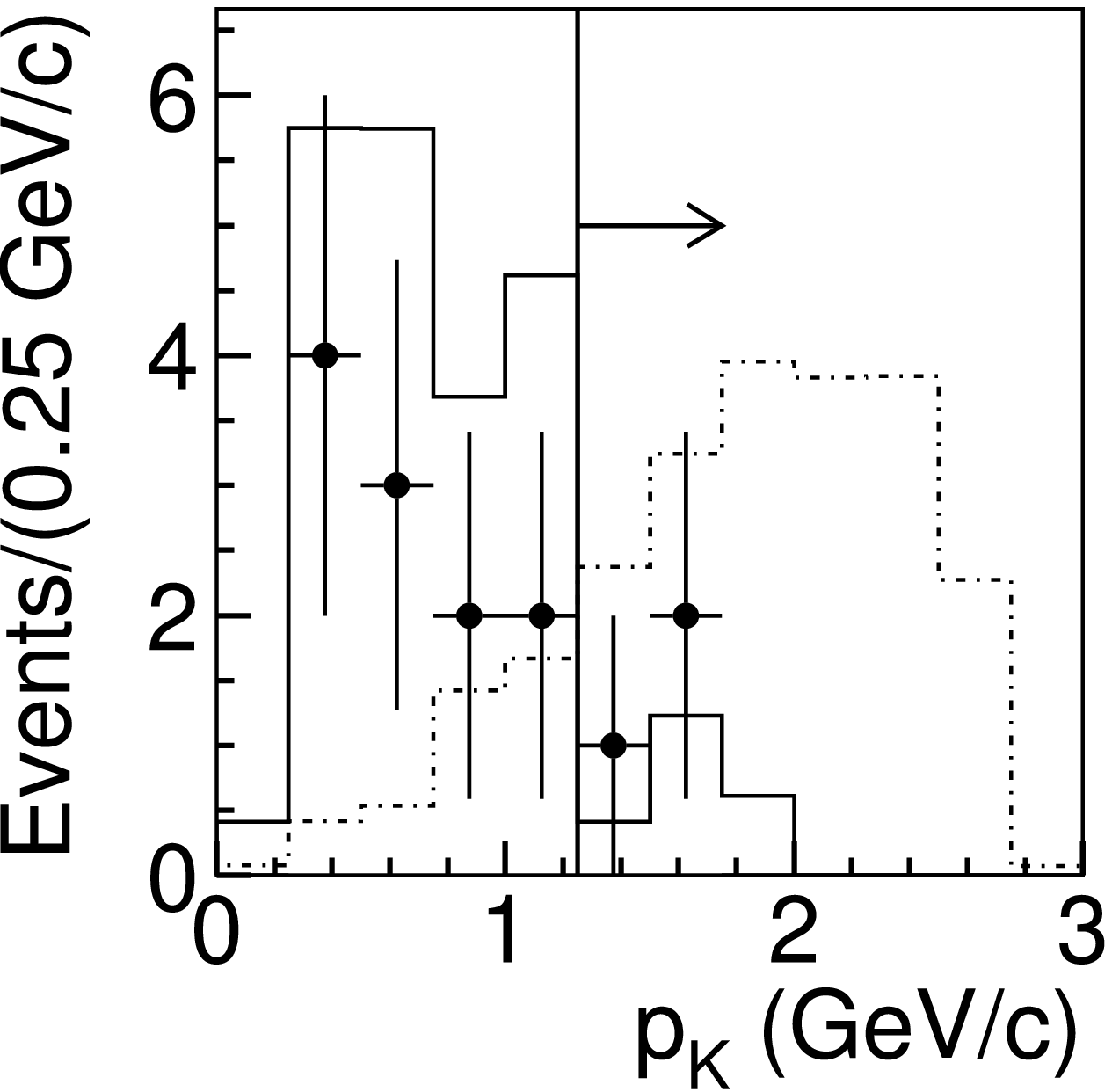}}
\parbox{4.0cm}
{\includegraphics[width=4.0cm,height=3.5cm]{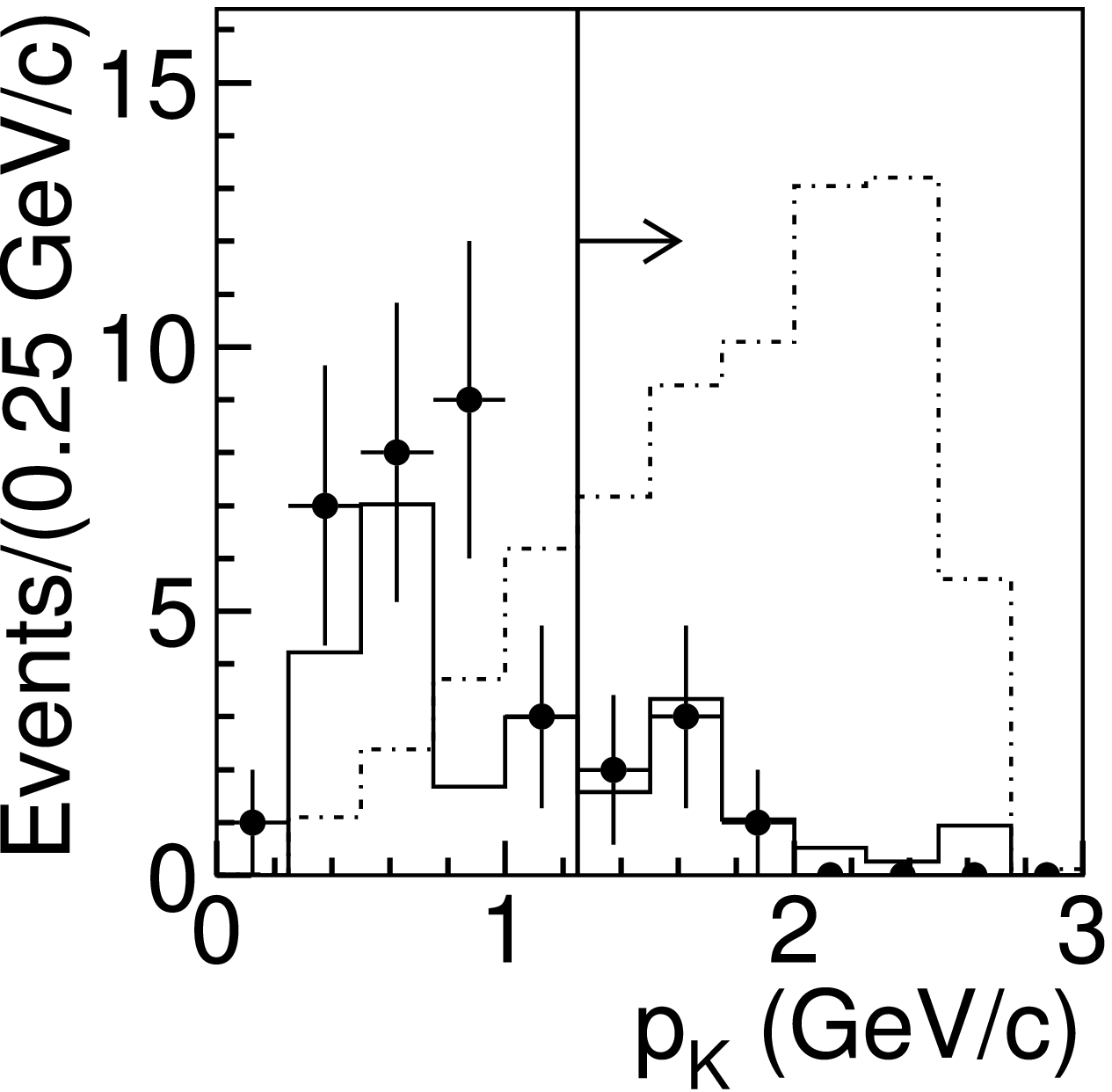}}
\caption{ The $p_{K}$ distribution for (a)
$\tagBhad$ and (b) $\tagBlep$ events after applying the full $\Bknunu$
 selection except for the $p_{K}>1.25$~GeV/$c$ requirement. 
The data and background MC samples are represented by the
points and solid histograms respectively.
The dotted line indicates the signal MC prediction for $\Bknunu$ with arbitrary normalization.}
\label{fig:pstar}
\end{center}
\end{figure}

In the $\tagBlep$ analysis, purely combinatorial backgrounds are estimated by 
examining sideband regions of the reconstructed
$\Dz$ invariant mass distribution, $m_{\Dz}^{\rm reco}$, defined by  
$3\sigma < |m_{\Dz}^{\rm reco} -m_{\Dz}| < 10\sigma$ as
 shown in ~Fig.~\ref{fig:mes2}a for the $\Dz \to \Km\pip$ mode.  
The sideband yields are scaled to the signal region under the assumption that 
the combinatorial component is flat throughout the $\Dz$ mass distribution.
This assumption has been validated using samples of events in which two
or three charged tracks not associated with the $\tagB$ 
reconstruction are present.
The total combinatorial background in the $\tagBlep$ analysis
 is estimated to be $N_{K}^{\rm bg} \geq 3.4\pm1.2$.
Although the peaking background prediction in the $\tagBlep$ analysis 
 have been studied in MC and are shown in 
Figs~\ref{fig:eextra} and~\ref{fig:pstar}, 
the peaking background in the final selection is not
subtracted.

In the $\tagBhad$ analysis, the combinatorial
background can be reliably estimated by extrapolating the observed yields 
in the $\mes$ sideband region into the $\mes$ signal region, indicated in  
Fig.~\ref{fig:mes2}b, yielding $2.0 \pm 0.7$ events.
The quoted uncertainty is dominated by the sideband data statistics, but includes 
also the uncertainty in the combinatorial background shape which is estimated
by varying the shape over a range of possible models. 
The peaking background in the $\tagBhad$ analysis consists only of  $B^+B^-$ events in which 
the $\tagBhad$ has been correctly reconstructed, and is estimated 
directly from $B^+B^-$ MC simulation.  MC yields are 
validated by direct comparison with data in samples of events in which the full signal selection is 
applied, except that either $\Eextra > 0.5$~GeV, or more than one charged 
track remains after the $\tagB$ reconstruction. 
Uncertainties in the peaking background are dominated by
the MC statistical uncertainty ($42\%$). Other systematic errors include
the overall $\tagB$ yield ($7\%$), 
the remaining charged track multiplicity ($5\%$),
the particle mis-identification rates for the $K^\pm$ selection ($6.3\%$), and the EMC energy modeling ($8\%$). 
The total peaking background in the $\tagBhad$ analysis is estimated to be
 $1.9 \pm 0.9$.  The total (combinatorial+peaking) background in the $\tagBhad$ analysis 
is estimated to be $N_{K}^{\rm bg}= 3.9\pm1.1$ events.

The $\Bknunu$ branching fraction is calculated from
\beq
\mathcal{B}(\Bknunu) = \frac{N_{K}^{\rm obs}-N_{K}^{\rm bg}}{N_{\Bpm}\cdot \eps_{K}} 
\eeq
where $N_{K}^{\rm obs}$ is the total number 
of observed events in the signal region.
$N_{\Bpm} = (88.9 \pm 1.0)\times 10^{6}$ is the estimated number of 
$\B^\pm$ mesons in the data sample and 
$\eps_{K}$ is the total efficiency. 
A total of $N_{K}^{\rm obs} = 6~(3)$ $\Bknunu$ candidate events are found in data in the $\tagBlep~(\tagBhad)$ analysis. 
The $p_{K}$ distributions for $\Bknunu$ signal 
events in the $\tagBlep$ and $\tagBhad$ analysis are shown in Fig.~\ref{fig:pstar}.

Branching fraction upper limits are computed using a modified 
frequentist approach, based on~\cite{ref:cousins}, which 
models systematic uncertainties using Gaussian distributions.  
For both the $\tagBlep$ and $\tagBhad$ searches, 
$\Bknunu$ limits are set 
at the branching fraction value at which it is estimated that $90\%$ of 
experiments would produce a yield which is greater than 
the number of signal events observed. 
Limits of
$\mathcal{B}(\Bknunu)_{\rm sl} < 7.0 \times 10^{-5}$ and 
$\mathcal{B}(\Bknunu)_{\rm had} < 6.7 \times 10^{-5}$ are 
obtained for the $\tagBlep$ and $\tagBhad$ searches respectively. 
Since the two tag $B$ samples are statistically independent, we can  
combine the results of the two analyses to derive a
limit of $\mathcal{B}(\Bknunu) < 5.2 \times 10^{-5}$ at the 
90$\%$ C.L. 

\begin{figure}[!htb]
\begin{center}
\parbox{4.0cm}
{\includegraphics[width=4.0cm,height=3.5cm]{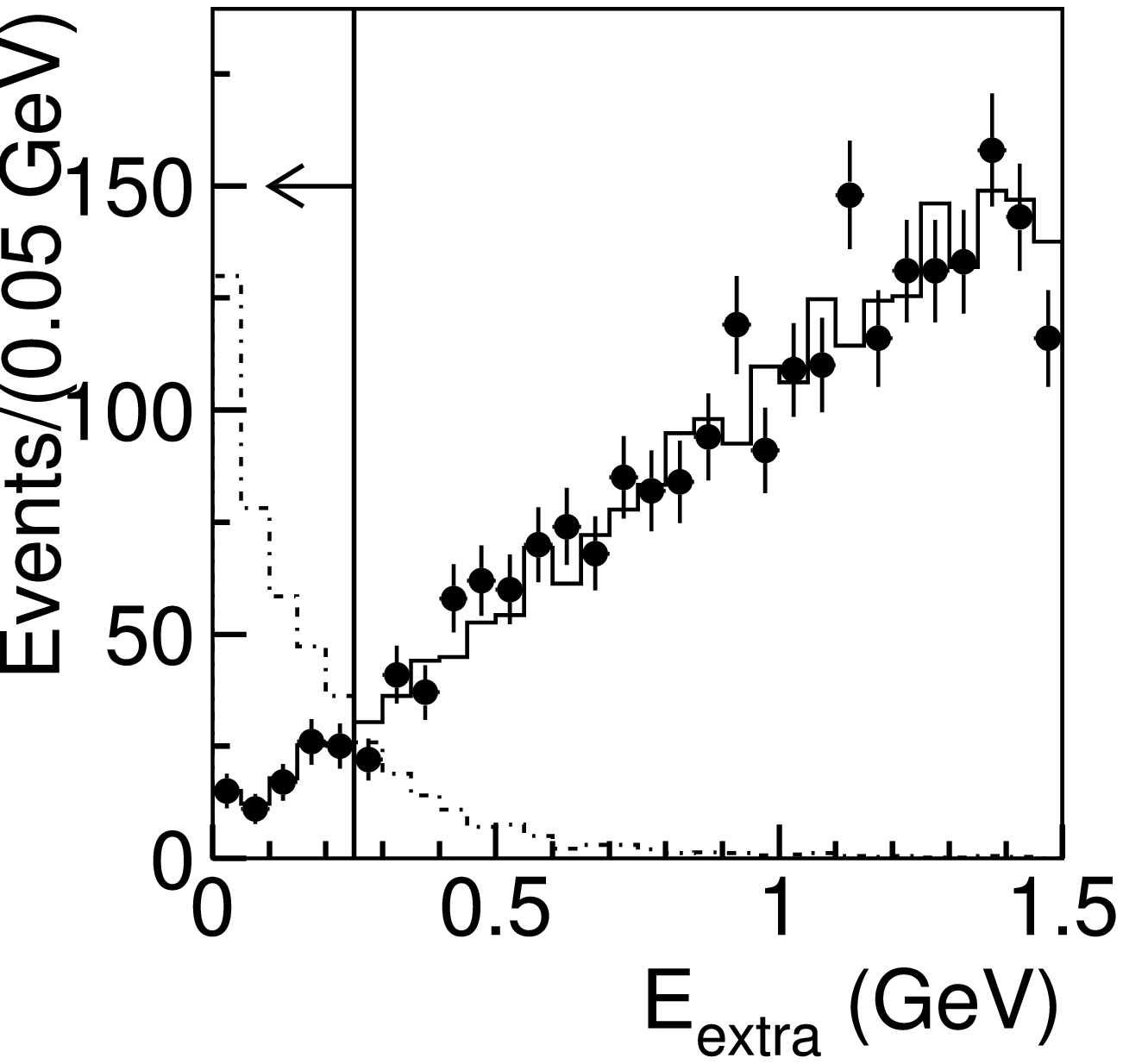}}
\parbox{4.0cm}
{\includegraphics[width=4.0cm,height=3.5cm]{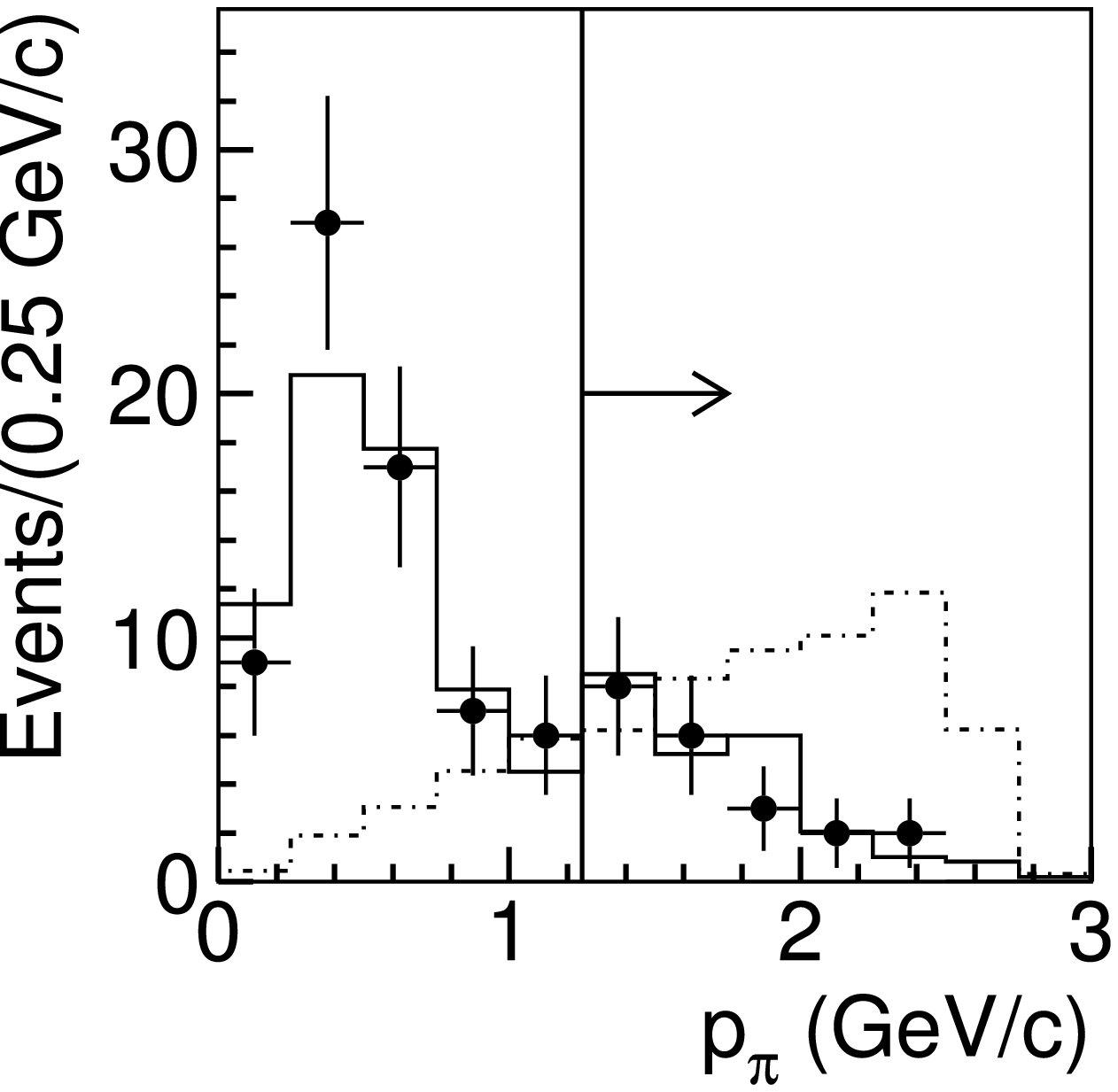}}
\caption{ The $\Eextra$ (a) and $p_{\pi}$ (b) distributions for $\Bpinunu$ in the $\tagBhad$ sample.
Events shown in the $\Eextra$ distribution are required to
have a reconstructed $\tagB$ and exactly one additional track satisfying the pion selection requirements. 
The $p_{\pi}$ distribution has all signal selection requirements applied other than the $p_{\pi}$ cut.  
The data and background MC samples are represented by the 
points and the solid histogram respectively. 
The dotted line indicates the signal MC prediction with arbitrary normalization.}
\label{fig:pinunu}
\end{center}
\end{figure}

We also report a limit on exclusive $\Bpinunu$ branching fraction using only the $\tagBhad$ sample.
The same methodology as for the $\Bknunu$ search is applied to the $\Bpinunu$ search except that
the single additional track is required not to satisfy either kaon or electron PID criteria.
The $\Eextra$ and $p_{\pi}$ distributions for $\Bpinunu$ are shown in Fig.~\ref{fig:pinunu}.
The overall $\Bpinunu$ selection efficiency is estimated to be $\pisigeffHad$, where the quoted uncertainties
include an estimated $2\%$ PID uncertainty, and other contributions to the systematic uncertainty are similar to $\Bknunu$.  
The peaking and non-peaking backgrounds are estimated to be $15.1 \pm 3.1$ events 
and $9.0\pm 1.8$ events respectively, with similar systematic uncertainties to the $\Bknunu$ analysis.
The search selects $N_{\pi}^{\rm obs} = 21$ candidates in data with an estimated total background of 
$N_{\pi}^{\rm bg}= 24.1 \pm 3.6$, resulting an 
upper limit of $\mathcal{B}(\Bpinunu)_{\rm had} < 1.0 \times 10^{-4}$ at the $90\%$ C.L..

We see no evidence for a signal in either of the reported decay modes. 
The $\mathcal{B}(\Bknunu)$ limit reported here is approximately one order 
of magnitude above the SM prediction and represents the most stringent 
experimental limit reported to date.


We are grateful for the excellent luminosity and machine conditions
provided by our \pep2\ colleagues, 
and for the substantial dedicated effort from
the computing organizations that support \babar.
The collaborating institutions wish to thank 
SLAC for its support and kind hospitality. 
This work is supported by
DOE
and NSF (USA),
NSERC (Canada),
IHEP (China),
CEA and
CNRS-IN2P3
(France),
BMBF and DFG
(Germany),
INFN (Italy),
FOM (The Netherlands),
NFR (Norway),
MIST (Russia), and
PPARC (United Kingdom). 
Individuals have received support from CONACyT (Mexico), A.~P.~Sloan Foundation, 
Research Corporation,
and Alexander von Humboldt Foundation.

\end{document}